\documentclass{article}

\usepackage{PRIMEarxiv}

\usepackage[utf8]{inputenc} % allow utf-8 input
\usepackage[T1]{fontenc}    % use 8-bit T1 fonts
\usepackage{hyperref}       % hyperlinks
\usepackage{url}            % simple URL typesetting
\usepackage{booktabs}       % professional-quality tables
\usepackage{amsfonts}       % blackboard math symbols
\usepackage{nicefrac}       % compact symbols for 1/2, etc.
\usepackage{microtype}      % microtypography
\usepackage{lipsum}
\usepackage{fancyhdr}       % header
\usepackage{graphicx} 
\usepackage{natbib} % graphics
\usepackage{multirow}
\graphicspath{{media/}}     % organize your images and other figures under media/ folder

\title{3D Skin Segmentation Methods in Medical Imaging:\\ A Comparison
\thanks{\textit{\underline{Citation}}: 
\textbf{Authors. Title. Pages.... DOI:000000/11111.}} 
}
\author{
  Martina Paccini, Giuseppe Patan\'e \\
  CNR - IMATI \\
  Genova\\
  \texttt{\{martina.paccini, giuseppe.patane\}@cnr.it} \\}

\begin{document}
\maketitle

\begin{abstract}
Automatic segmentation of anatomical structures is critical in medical image analysis, aiding diagnostics and treatment planning. Skin segmentation plays a key role in registering and visualising multimodal imaging data. 3D skin segmentation enables applications in personalised medicine, surgical planning, and remote monitoring, offering realistic patient models for treatment simulation, procedural visualisation, and continuous condition tracking. This paper analyses and compares algorithmic and AI-driven skin segmentation approaches, emphasising key factors to consider when selecting a strategy based on data availability and application requirements. We evaluate an iterative region-growing algorithm and the TotalSegmentator, a deep learning-based approach, across different imaging modalities and anatomical regions. Our tests show that AI segmentation excels in automation but struggles with MRI due to its CT-based training, while the graphics-based method performs better for MRIs but introduces more noise. AI-driven segmentation also automates patient bed removal in CT, whereas the graphics-based method requires manual intervention. 
\end{abstract}

% keywords can be removed

\section{Introduction\label{intro}}
Automatic segmentation of anatomical components is widely studied in medical image analysis, focusing on identifying the boundaries of structures like organs, tissues, or pathological lesions for diagnostic or monitoring purposes~\citep{lenchik_automated_2019}. In particular, skin segmentation has a crucial role in registering and visualising heterogeneous data from different imaging modalities. \emph{Skin segmentation} refers to identifying and isolating skin regions in MR or CT images. It separates the skin from the background and internal anatomy for several medical purposes, such as registration and visualisation of heterogeneous data from different imaging modalities, surgery and simulation. More precisely, 3D skin segmentation provides a realistic and interactive 3D model of the patient’s skin surface, which can be used for (i) \emph{personalised medicine}, where healthcare providers can simulate different treatments on 3D models; (ii) \emph{surgery planning}, where 3D skin models are used to practice or visualise procedures before performing them on the patient, comparing pre- and post-treatment, and improving diagnosis accuracy; and (iii) \emph{remote monitoring}, where a digital twin can be updated with real-time data (e.g., changes in skin lesions or wounds), enabling doctors to track a patient’s condition remotely over time.

This paper analyses and compares previous skin surface segmentation works using algorithmic and AI-driven approaches (Sect.~\ref{overview}). We select a representative method from each category and examine the trade-offs between these approaches (Sect.~\ref{comparison}). Additionally, we conduct a comparative analysis across different imaging modalities and anatomical regions, evaluating their performance in terms of accuracy, computational cost, system requirements, and limitations. Specifically, we examine a graphics-based segmentation~\citep{paccini2024us}, based on an iterative region-growing algorithm, and TotalSegmentator~\citep{wasserthal2023totalsegmentator}, based on deep learning. To quantify segmentation differences, we use the Hausdorff distance, complemented by a visual comparison of segmentation results from abdominal MRI, head CT, and total-body CT, as well as an evaluation on a dataset of MRIs and CTs capturing different breathing phases~\citep{data/ICSFUS_2022,bauer2021generation}. Our findings indicate that while AI segmentation excels in automation, it struggles with MRI segmentation due to its training in CT data. Conversely, the graphics-based method performs better for MRIs but introduces more noise. Additionally, AI-driven segmentation automatically removes patient beds in CT scans, whereas the graphics-based method requires manual adjustments (Sect.~\ref{sec:CONCLUSION}).% {\color{red}{Finally, we extend the segmentation discrepancies from the skin surface to the inner volume, generating a volumetric uncertainty map. This map is valuable for medical decision-making, as uncertainty visualisation is fundamental in medical imaging~\citep{gillmann2021uncertainty}.}}

\section{Skin segmentation methods: an overview\label{overview}}
Although skin segmentation is relevant for various medical applications, it has typically been considered a preprocessing step or an auxiliary task within broader imaging and analysis frameworks. With the diffusion of 3D depth cameras and other 3D scanning systems, skin segmentation has gained increased importance. From the perspective of medical applications, it is preferable to perform skin segmentation across different types of volumetric anatomical images to enhance accuracy and facilitate the integration of multimodal data. Skin segmentation in MRI is less common than in CT or dermatological imaging, as MRI primarily focuses on internal structures such as the brain, muscles, and organs. Additionally, its lower resolution for superficial structures compared with optical imaging limits its widespread use for skin analysis. However, MRI-based skin segmentation has important applications, particularly when high-resolution imaging of the skin and adjacent structures is required. 3D skin segmentation enables precise extraction of the skin surface while providing detailed visualisation of subcutaneous structures and surrounding tissues. Furthermore, MRI is highly effective in analysing deeper skin layers, connective tissue, and muscle, making it valuable in oncology, reconstructive surgery, and tissue engineering~\citep{scorza2021surgical,butz2000pre}.

\subsection{Algoritmic skin segmentation}
In \emph{manual segmentation}~\citep{rosset2004osirix,schindelin2012fiji,wolf2005medical,amira,yushkevich2006user}, radiologists or medical professionals manually, or semi-automatically, delineate the skin boundaries layer by layer in 3D MR images. Manual segmentation achieves a high accuracy and is used when other methods might miss subtle anatomical details; however, it is time-consuming and labour-intensive, especially for complex 3D datasets, and subject to inter-operator variability. \emph{Thresholding-based methods}~\citep{otsu1979threshold} segment the skin by selecting a specific intensity range in the MR image that corresponds to skin tissues, separating it from other structures. This approach is simple to implement and can work well when there is a clear contrast between skin and surrounding tissues, such as in CT images. However, MR intensity values can vary across scans, making it difficult to apply thresholding universally without adjustments. It also struggles in areas with low contrast between skin and adjacent tissues. Some works applied thresholding to skin identification on classical CT and MRI followed by morphological filters~\citep{jermyn_fast_2013} or 3D vector-based connected component algorithm~\citep{wang_fully_2012}. Thresholds have also been leveraged in other districts on the raw image~\citep{baum_does_2012} or after a pre-processing aimed at edge enhancement~\citep{khang2022computer}. These works apply the thresholding on the whole image to classify the pixel in the background (black) or body (white) and then use other filtering methods to clean the obtained result.

\emph{Region-growing algorithms}~\citep{pan2007bayes} start from a "seed point" (a pixel known to be part of the skin) and grow outward, including neighbouring pixels based on similar intensity or texture features. Region growing allows for the segmentation of contiguous areas, making it useful for irregular skin surfaces or areas with varying thicknesses; however, these algorithms are sensitive to noise and may require manual intervention to correct inaccuracies. It can sometimes leak into non-skin areas if the intensity contrast is weak.  In~\citep{bosnjak20073d}, the region growing algorithm is applied to identify the skin surrounding the skull in surgical applications from MRI volumes. The seed is manually positioned by an operator, resulting in segmentation outcomes that include internal brain structures. While this may be advantageous depending on the application, it poses challenges when the objective is to isolate only the skin. In~\citep{caballo2018unsupervised}, breast skin segmentation is achieved by a region-growing which uses constraints from the previously extracted skin centerline to add robustness to the model and to reduce the false positive rate. An energy-minimizing active contour model is then applied to classify adipose tissue voxels by including gradient flow and region-based features. This method works for CT images, where the background is distinctly separated from the skin.~\citep{teuwen2021deep} leveraged the high contrast of skin in \emph{Digital Breast Tomosynthesis} images to implement a fast-seeded region-growing algorithm. The primary objective was to differentiate the outer skin layer from the adjacent internal layers. The algorithm initiates the segmentation by placing a subset of seeds along the outer edge of the skin layer. It subsequently expands the segmented region by incorporating voxels whose intensity values were greater than or equal to the mean intensity of the seed points.

\emph{Edge detection and gradient-based methods}~\citep{jing2022recent} detect sharp changes in image intensity, which typically correspond to the boundaries between the skin and other tissues. These methods are effective in capturing clear skin boundaries, especially in high-contrast regions of the image; however, MRI images often have blurred boundaries or low contrast between skin and adjacent tissue, making it difficult for edge detection to perform consistently, and post-processing is often needed. In~\citep{beare_automated_2016}, the Watershed transform from markers is used on a gradient image containing light to do dark transitions obtained from T1-MRI. On 3T MRI with T2-weighted sequence, a combination of Canny filter, selection of boundaries and a local regression are applied to delimit the different skin layers~\citep{ognard_edge_2019}. However, 3T MRIs still need to be diffused as low-field MRIs in everyday clinical practice, which makes this method hard to be applied to new portable systems. \emph{Active contours/snakes}~\citep{chan2001active} are boundary-based methods where an initial contour is placed around the skin boundary and iteratively adjusts itself based on internal (smoothness) and external (image intensity gradients) forces. Even though these methods are effective in segmenting smooth and continuous boundaries like the skin surface, they are generally susceptible to noise and may struggle to accurately follow boundaries in low-contrast areas of the MRI scan and are usually focused on a specific image acquisition system when dealing with skin segmentation~\citep{lee_automated_2018,le2016volume}.

\emph{Atlas-based segmentation}~\citep{hirsch2021segmentation} applies predefined anatomical atlases (i.e., standard anatomical templates), which are registered and deformed to fit the patient's MRI data and contain prior information about the location and structure of the skin. On the one hand, these methods help to automate the segmentation process using prior knowledge of human anatomy and can improve accuracy in complex cases where the skin boundaries are not obvious. On the other hand, these methods require high-quality registration of the atlas to the patient’s MRI, which can be difficult if there are anatomical variations or large deformities. \emph{Level-set methods}~\citep{al2014breast} evolve a contour over time based on internal and external forces that help it conform to the boundaries of the skin. It works by minimising a cost function that typically includes terms for image gradients and smoothness constraints. These methods are particularly useful for segmenting irregularly shaped or complex skin surfaces; however, they are computationally intensive and may require fine-tuning of parameters for each case.

The \emph{graphics-based segmentation}~\citep{paccini2024us} begins from a background pixel, typically in the corners, assuming the subject is centred. This pixel’s intensity is checked against the skin isovalue (automatically computed, refer to Sect.~\ref{Res_qualitative}) to confirm its classification as background. If valid, then the corresponding grid element is set to 0, and the pixel is marked as visited. To propagate segmentation, a list tracks pixels yet to be evaluated, initially containing the neighbours of the starting pixel. If a pixel’s intensity exceeds the isovalue, indicating the body edge, then its grid value is set to 1, and no further neighbours are added. The process continues until all relevant pixels are classified. By the end, the grid contains 0s for the background, 1s at the skin boundary, and the \textit{initial value} (e.g., 2) inside the body. This method is applied to all slices in the volume, with computational cost scaling linearly with the number of voxels. While this method shares conceptual similarities with the region-growing algorithms~\citep{teuwen2021deep}, it differs in requiring a single seed and applying it to other imaging modalities.

\subsection{AI-driven skin segmentation}
The advancements in deep learning methods have spread them within different branches of the medical imaging domain. 
\emph{Machine learning and deep learning approaches}, particularly deep learning algorithms like CNNs, are trained on large datasets of labelled MRI scans to automatically segment the skin~\citep{gillot2022automatic}. For instance, 3D CNNs work directly with 3D MRI data to segment the skin volumetrically, learning to identify skin features across slices. Regarding 3D medical image segmentation, current AI approaches are typically designed for specific modalities or disease types and often struggle to generalise across the wide range of medical image segmentation tasks. Among the most prominent recent segmentation techniques, nnUNet~\citep{isensee2021nnu} is a semantic approach that automatically adapts to the characteristics of a given dataset. It is based on supervised learning, meaning that specific training data for the target application must be provided. nnUNet has been successfully applied to skin segmentation in~\citep{bi2024open}, particularly in head MRIs for surgical guidance purposes.

With the introduction of foundation models, the possibility to generalise among different imaging modalities of deep learning methods has increased. A notable example is medSAM~\citep{ma2024segment}, a foundation model designed to bridge the AI segmentation gap by enabling universal medical image segmentation. The model is developed on a large-scale medical image dataset with 1,570,263 image-mask pairs, covering 10 imaging modalities and over 30 cancer types. The authors conduct a comprehensive evaluation of 86 internal validation tasks and 60 external validation tasks, demonstrating better accuracy and robustness than modality-wise specialist models. However, it does not provide the skin segmentation among the regions considered.

The \emph{Softmax for Arbitrary Label Trees} (SALT)~\citep{koitka2024salt} segments CT images by leveraging conditional probabilities to map the hierarchical structure of anatomical landmarks. The model was developed using the SAROS dataset from the Cancer Imaging Archive, comprising 900 body region segmentations from 883 patients~\citep{koitka2024saros}. The dataset was further enhanced by generating additional segmentations with the \emph{TotalSegmentator}~\citep{wasserthal2023totalsegmentator}, for a total of 113 labels. The model was trained on 600 scans, while validation and testing were conducted on 150 CT scans. SALT used the hierarchical structures inherent in the human body to achieve whole-body segmentations with an average of 35 seconds per CT scan. However, among the labels that the SALT framework can identify, there is not the explicit identification of the skin. Finally, MONAI Auto3DSeg represents an extension for 3D Slicer (marchio) for fully automatic AI segmentation of images. MONAI is a freely available, community-supported, and consortium-led PyTorch-based framework for deep learning in healthcare. MONAI extends PyTorch to support medical data, with a particular focus on imaging and provides purpose-specific AI model architectures, transformations and utilities that streamline the development and deployment of medical AI models~\citep{DIAZPINTO2024103207}. The possibility to segment the body from CT as abdominal, chest, or whole-body CT is trained using the TotalSegmentator models.

The TotalSegmentator, includes both the trained network able to segment the CT images and a large dataset with labels.  Among the various tasks this model can accomplish, there is also the body segmentation, which provides the body region, trunk, extremities and skin segmentation. The model's applications include surgery, dosimetry, research, age-dependent value provision, and biomarker identification, demonstrating versatility in various medical scenarios.

\section{Skin segmentation comparison\label{comparison}}
Among all the AI segmentation methods previously revised, the TotalSegmentator is particularly interesting since it allows the explicit segmentation of the body edge and external surface and is ready to use without further training datasets. Other approaches instead do not focus on this specific task but provide only the identification of the background. Given the great relevance of the TotalSegmentator both as a dataset to train new models and the actual network to perform the segmentation, we consider it as the representative work for the AI skin segmentation method in the comparison. Regarding the algorithmic methods for skin segmentation, we consider the graphics-based~\citep{paccini2024us} for its generality with respect to the image modality and anatomical district.

\subsection{Algorithmic vs. AI-based skin segmentation: a comparative analysis}
The choice between algorithmic segmentation (e.g., graphic-based) and AI-driven approaches (e.g., TotalSegmentator) depends on several factors, including the specific requirements of the application, the availability of annotated data, and the computational resources at disposal~\citep{tripathi2025disruptive}. Each method has unique strengths and limitations, making them suitable for different medical imaging applications. 

Algorithmic segmentation methods, including region-growing, thresholding, and graph-based techniques, provide a rule-based and interpretable framework for extracting anatomical surfaces~\citep{tripathi2025disruptive}. These methods are widely used due to their independence from training data and their ability to work across multiple imaging modalities, including MRI and CT. Their transparency allows users to adjust parameters based on anatomical knowledge, making them particularly valuable when handling diverse datasets. However, algorithmic segmentation methods face challenges in generalizability and adaptability. They often require extensive parameter tuning and manual intervention to account for variations in image quality, contrast, and anatomical structures. While computationally efficient, these techniques may struggle with complex or low-contrast structures, particularly in MRI, where skin segmentation is less common due to the lower resolution of superficial structures compared to optical imaging~\citep{teng2024literature}.
\begin{figure}[t]
	\centering
	\begin{tabular}{cccc}
		\multicolumn{2}{c}{Graphics-based segmentation} &\multicolumn{2}{c}{TotalSegmentator}\\
		\includegraphics[width=0.20\linewidth]{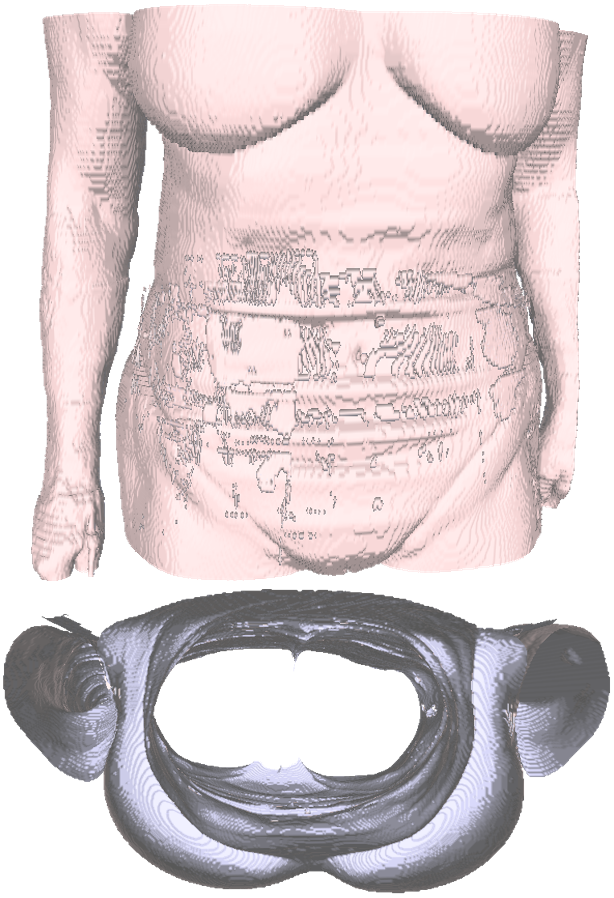}&
		\includegraphics[width=0.20 \linewidth]{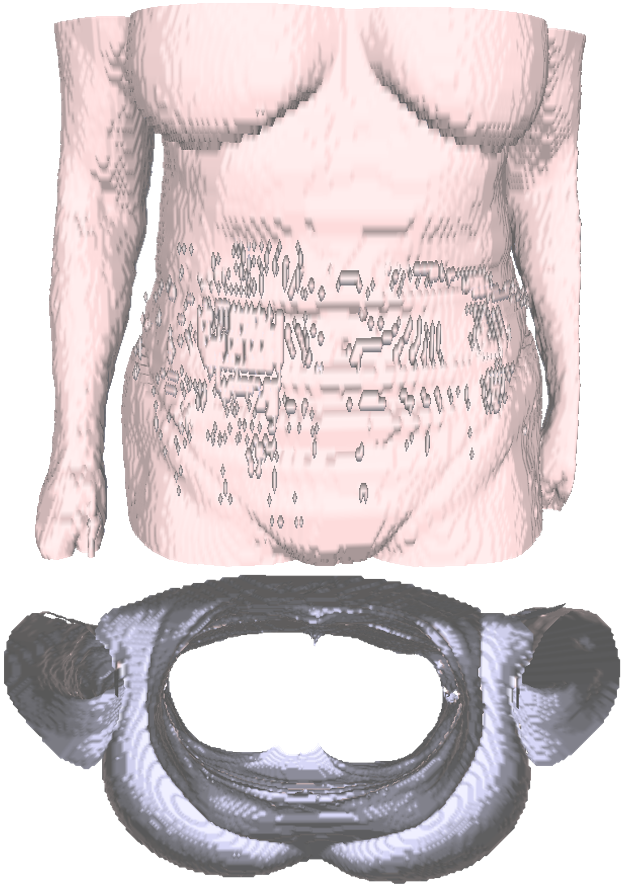}&
		\includegraphics[width=0.20 \linewidth]{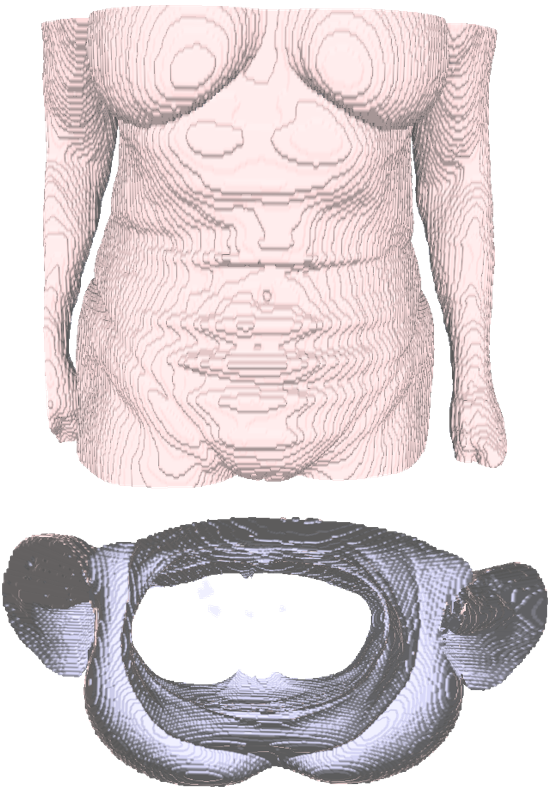}&
		\includegraphics[width=0.20 \linewidth]{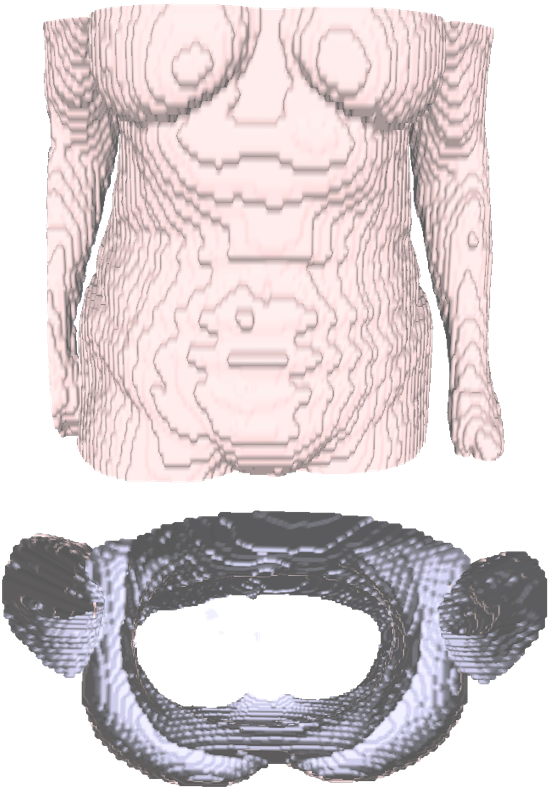}
	\end{tabular}
	\caption{\label{fig:subsample} Skin surfaces segmented from a high (a-b, left) and low resolution (a-b, right) image. According to the Hausdorff distance between the high and low-resolution segmentations, the graphics-based method has a deviation (33.9 mm) that is lower than (56.5 mm) the TotalSegmentator.}
\end{figure}

Deep learning-based segmentation tools provide a high degree of automation, making them well-suited for large-scale clinical applications. Trained on extensive labelled datasets, these models can achieve high segmentation accuracy, particularly in structured imaging modalities like CT. AI methods excel at capturing fine details, reducing the need for manual intervention, and streamlining workflows. Despite these advantages, AI-driven segmentation has limitations. The imaging modality dependence reduces its flexibility across different imaging techniques, and the creation of a large annotated dataset presents considerable difficulties. Furthermore, AI models are computationally demanding, requiring high-performance hardware for both training and inference. The “black-box” nature of deep learning also poses challenges in clinical settings, where interpretability is critical for trust and validation~\citep{teng2024literature,tripathi2025disruptive,kamel2021digital}. Additionally, AI-based segmentation automatically removes the patient’s bed in CT scans, which can be beneficial for analysis but may introduce inconsistencies when comparing across different modalities or datasets. Unlike algorithmic methods, which can be fine-tuned manually, AI-based models require extensively annotated datasets to adapt to new imaging conditions.

When selecting a segmentation approach for skin segmentation between algorithmic methods and AI-driven techniques, it is important to consider the following trade-offs. \emph{Multi-modality support}: algorithmic segmentation is effective across MRI and CT, whereas AI-based is optimised primarily for CT and struggles with MRI if not properly trained; \emph{Automation vs. control}: deep learning methods provide full automation, reducing user input, while algorithmic segmentation allows manual refinement and customisation; \emph{Computational cost}: algorithmic methods are computationally lightweight and run on standard hardware, whereas AI-based segmentation requires GPUs and high-performance computing resources; \emph{Accuracy~$\&$ robustness}: AI-based segmentation excels in structured modalities like CT but may require retraining for other modalities. Algorithmic methods, while more adaptable, require careful parameter tuning; \emph{Interpretability}: algorithmic approaches are transparent and explainable, whereas deep learning models function as black boxes with limited user control.

%
%\begin{figure}[t]
%	\centering
%	\begin{tabular}{cccc}
	%		(a)\includegraphics[width=0.23 \linewidth, height=0.30 \linewidth]{US-COREG-images/CTheadAI.png}&
	%		(b)\includegraphics[width=0.23 \linewidth, height=0.30 \linewidth]{US-COREG-images/CTheadBG.png}&
	%		(c)\includegraphics[width=0.23 \linewidth, height=0.30 \linewidth]{US-COREG-images/HousdorffCThead.png}&
	%		\includegraphics[width=0.15 \linewidth, height=0.30 \linewidth]{US-COREG-images/colorbar.png}
	%	\end{tabular}
%	\caption{\label{fig:CTheadcomp}Skin segmentation from head CT using the (a) TotalSegmentator and (b) the graphics-based segmentation, (c) distance distribution (in mm) of surface (a) from (b), where the main difference is the head holder in the graphics-based segmentation.} 
%\end{figure}
%
\paragraph*{Application to skin segmentation in MRI and CT}
Skin segmentation plays a crucial role in medical imaging applications such as surgical planning, image registration, and patient-specific modelling. Algorithmic skin segmentation methods, like the graphics-based, are generally modality-agnostic, making them applicable to both MRI and CT without retraining and with flexibility in adjusting parameters for different imaging conditions. These methods enable interpretable segmentation and can be refined using domain knowledge. However, algorithmic methods are sensitive to image quality, scanner variations, and parameter selection. The leverage of threshold values struggles with intensity variations, and the region-growing approach can lead to segmentation leakage if the contrast between skin and surrounding tissues is poor. 

The TotalSegmentator provides consistent and efficient skin extraction from CT images. Its deep learning-based architecture allows it to handle complex anatomical structures without manual parameter tuning, making it highly effective for large-scale applications. However, TotalSegmentator’s primary limitation is its reliance on CT training data, which affects its performance on MRI. Unlike algorithmic methods, which can be manually adjusted to different modalities, TotalSegmentator requires retraining with MRI-specific data to achieve comparable accuracy. Additionally, AI-based methods demand substantial computational resources, making them less accessible in resource-constrained environments.

\subsection{Qualitative and quantitative analysis\label{Res_qualitative}}
Testing the graphics-based segmentation on sub-sampled images shows that its accuracy is maintained, with only a reduction in resolution (Fig.~\ref{fig:subsample}). Experiments with T2 MR and CT images show that normalising intensities between 0 and 1 allow a skin isovalue of 0.1 to achieve precise segmentation. For fully automated T2 MRI segmentation, computing the gradient image and using an isovalue of 0.01 yielded optimal results. Fig.~\ref{fig:subsample} shows the Totasegmentator results obtained with the same sub-sampling performed on the graphics-based method test. The image's resolution, even in the original case, is lower than the graphics-based one because the TotalSegmentator already performs a sub-sampling on the input image.

We compare the two skin segmentation methods on different imaging modalities and anatomical districts. Representing the segmented skin surfaces as meshes extracted by the Marching Cubes~\citep{lorensen1998marching}, their differences are evaluated by computing their Hausdorff distance: the higher the value of Hausdorff distance, the more the results differ. Calling the graphics-based segmented surface~$\mathbf{X}_{1}$ and the TotalSegmentator surface~$\mathbf{X}_{2}$ we compute the Hausdorff distance as the local distribution of the minimum distance of each vertex of the surface at~$\mathbf{X}_{2}$ from the vertices of the surface at~$\mathbf{X}_{1}$:~${d_{\mathbf{X}_{1}}\left(\mathbf{X}_{2}\right):=\max _{\mathbf{x}\in \mathbf{X}_{1}}\left\{ \min _{\mathbf{y}\in \mathbf{X}_{2}}\left\{ \left\|\mathbf{x}-\mathbf{y}\right\| _{2}\right\} \right\}}$, where~$\mathbf{X}_{1}$ and~$\mathbf{X}_{2}$ represent the skin vertices coordinates of the surfaces mesh.
\begin{figure}[t]
	\centering
	\begin{tabular}{cccc} % L'ultima colonna separata con "|" per la colorbar	
		(a)\includegraphics[width=0.2\linewidth]{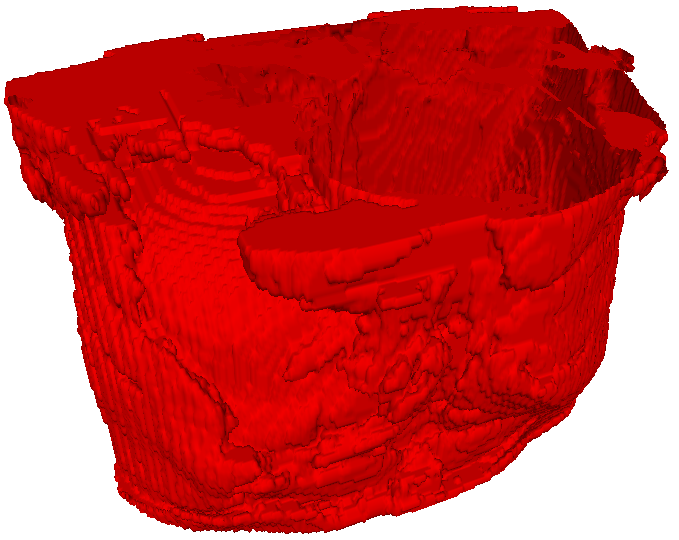}&
		\includegraphics[width=0.2\linewidth]{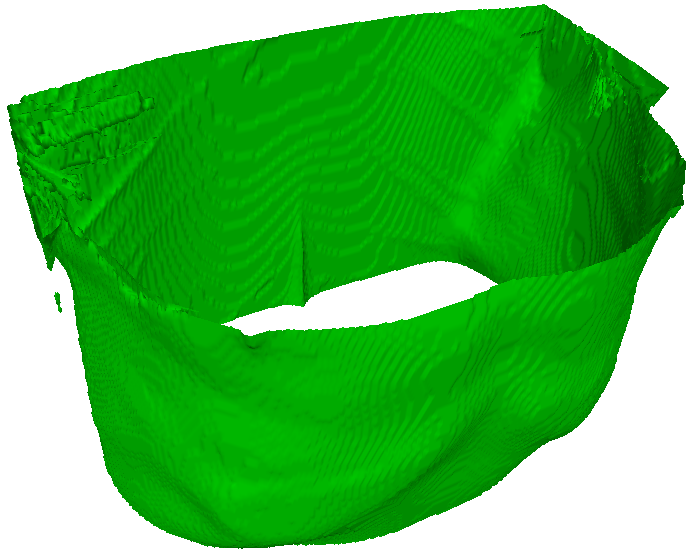}&
		\includegraphics[width=0.2\linewidth]{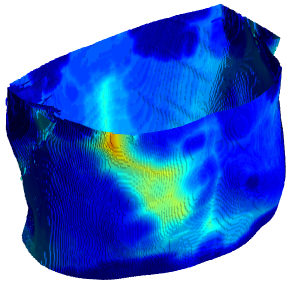}& 
		\multirow{3}{*}[1in]{\includegraphics[width=0.2\linewidth, height=0.53 \linewidth]{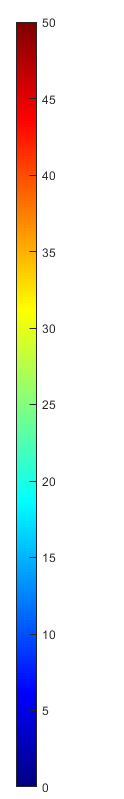}} \\
		(b)\includegraphics[height=0.3\linewidth]{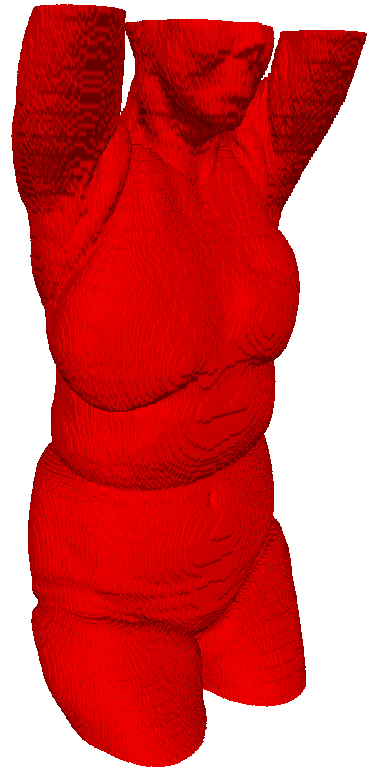}&
		\includegraphics[height=0.3\linewidth]{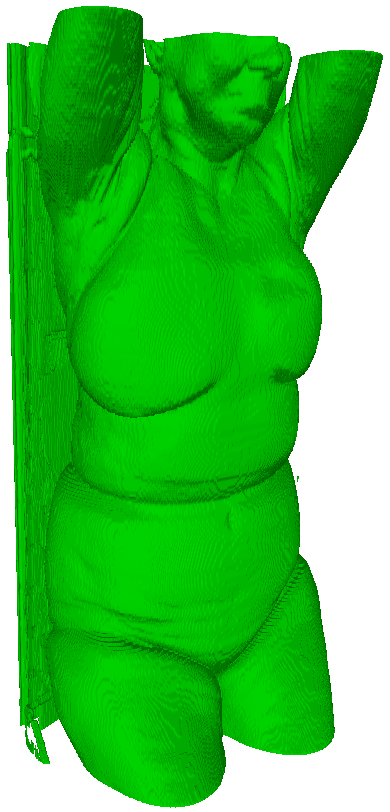}&
		\includegraphics[height=0.3\linewidth]{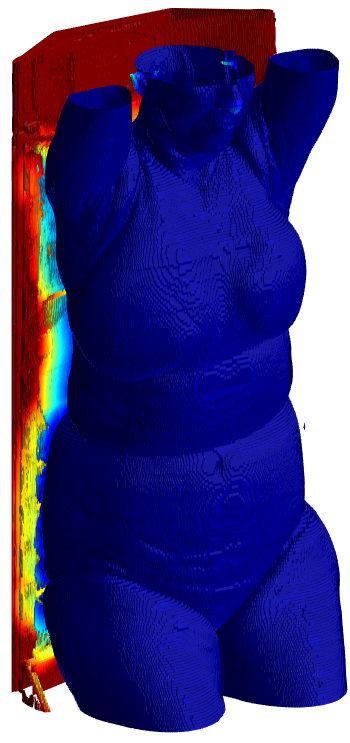}& \\
	\end{tabular}
	\caption{\label{fig:MRIabdcomp}Skin segmentation from an abdominal MRI based on (red) the TotalSegmentator and (green) the graphics-based segmentation and distance distribution between the two surfaces. (a) Abdominal MRI. (b) Total body CT. The color bar unit measure is mm.}
\end{figure}

%
%\begin{figure}[t]
% \centering
%	\begin{tabular}{cccc}
	%		(a)\includegraphics[height=0.35 \linewidth]{US-COREG-images/CTbodyAI.png}&
	%		(b)\includegraphics[height=0.35 \linewidth]{US-COREG-images/CTbodyGB.png}&
	%		(c)\includegraphics[height=0.35 \linewidth]{US-COREG-images/HousdorffBody.png}&
	%		\includegraphics[width=0.15 \linewidth, height=0.35 \linewidth]{US-COREG-images/colorbar.png}
	%	\end{tabular}
%	\caption{\label{fig:CTbodycomp} Skin segmentation from a total-body CT based on (a) the TotalSegmentator and (b) the graphics-based segmentation: (c) distance distribution of surface (a) from {\color{red}{good segmentation}}. The Hausdorff distance between the two surfaces is  H=156.707mm.}
%\end{figure}

Fig.~\ref{fig:MRIabdcomp}(a) shows the results obtained segmenting an abdomen MRI. Graphics-based segmentation provides a better result than the TotalSegmentator, as the latter needs proper training to segment a specific typology of images. In this case, the AI network was trained on CT and not MRIs, obtaining a worse result. The existence of a high number of MRI sequences makes this type of medical image particularly complex to be segmented from AI without biases since a large number of images is required to train the network for each different sequence. In the MRI case, the evaluation of the distribution of distances between the two skins shows a high level of difference between the two in the regions where the AI method could not follow the body boundaries in the original image. Moreover, even with CT images the TotalSegmentator's surface could present holes caused by the edges of the image, while in graphics-based segmentation, the padding prevents the holes .
\begin{figure}[t]
	\centering
	\begin{tabular}{cccc}
		\includegraphics[height=0.35 \linewidth]{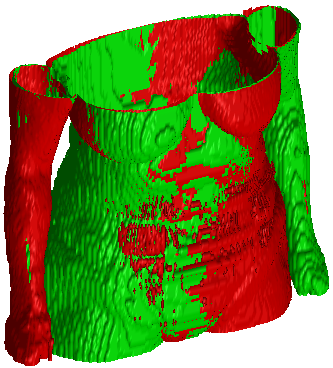}&
		\includegraphics[height=0.35 \linewidth]{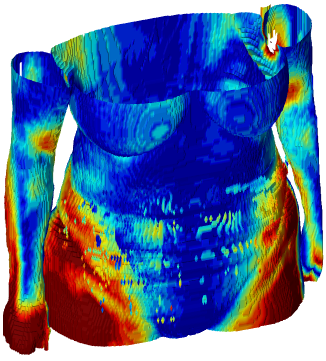}&	
		\includegraphics[height=0.35 \linewidth]{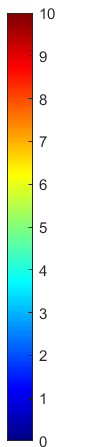}&
	\end{tabular}
	\caption{\label{fig:subcomparison}Comparison of the segmentations on a subsampled image. (a) Super-imposition of the two surfaces, (b) distance distribution (in mm).}
\end{figure}

Fig.~\ref{fig:MRIabdcomp}(b) shows the results on total body CT. In this case, both methods provide a good and clean segmentation, extracting solely the skin contour without internal organs or noise. However, the patient’s bed is still present in the graphics-based segmentation. The TotalSegmentator, instead, as for the head CT, can automatically remove the bed. Moreover, the TotalSegmentator's surface could present holes caused by the edges of the image, while in graphics-based segmentation, the padding prevents the holes. Table~\ref{tab:Hcomp} shows the results on the Hausdorff distance obtained comparing the surfaces extracted from the graphics-based segmentation and the TotalSegmentator. The distances reported in the table do not consider the presence of the bed or the head support to concentrate only on the possible differences between the skin surfaces. The values of Hausdorff distances for the different MRIs can be related to the difficulties of the TotalSegmentator with this image typology. The head MRI presents the lower value of Hausdorff distance since the TotalSegmentator can identify a clear distinction between the body and the background. However, the segmentation seems to be related more to the skull than the actual skin. The high distance value in the head CT is because the TotalSegmentator does not identify clear borders on all the image slices (which also comprehends the part of the chest). Still, it stops at the end of the head and does not segment the skin at the chest level. We compare the results obtained with the subsampled image of both methods and present the results in Fig.~\ref{fig:subcomparison}. The two surfaces are similar, but the higher difference is around the extremities.
\begin{figure}[t]
	\centering
	\begin{tabular}{cccc}
		\includegraphics[height=0.35 \linewidth]{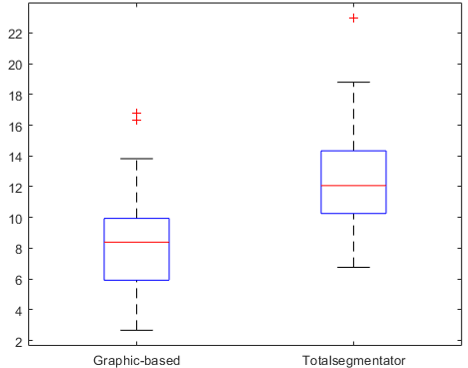}&
		\includegraphics[height=0.35 \linewidth]{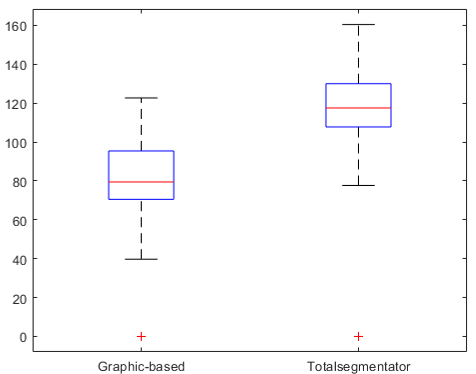}&	
	\end{tabular}
	\caption{\label{fig:confrontoDistribuzioni}Comparison of the mean distance (a) and Hausdorff distance (b) distribution on the whole dataset for the two segmentation methods considered. The value on the vertical axis indicates the distances in mm. The high value of distance is due to the presence of the arm in the MRIs, which were not present in the CT images.}
\end{figure}

The computational cost of the graphics-based segmentation is linear in the number of voxels and, thus, can run without any particular machine requirements. A sub-sampling of the original image is applied to reduce the segmentation time further. The TotalSegmentator runs on the CPU if it does not detect any GPU. In this way, the segmentation can be obtained also on an average machine even if the time required for the final results is extended. Running the TotalSegmentator on a GPU could accelerate the process. The method also provides an option for a faster segmentation that will run a lower resolution model (3mm instead of 1.5mm) for faster runtime and fewer memory requirements.
\begin{table}[t]
	\centering
	\caption{Comparison of the skin segmentations according to their Hausdorff distances. The evaluation considers different anatomical districts and acquisition methodologies. The Hausdorff distances are computed without considering beds or supports for the patient.}\label{tab:Hcomp}
	{\small{	\begin{tabular}{|l|l|l|} 
				\hline
				\textbf{Image} &\textbf{Anatomical} &\textbf{Hausdorff}\\
				\textbf{modality} &\textbf{district} &\textbf{distance (mm)}\\
				\hline
				MRI & Hand & 80.9\\ 
				\hline
				MRI T1& Abdomen & 110.14\\
				\hline
				MRI T2 & Abdomen & 103.9\\
				\hline
				CT & Whole body & 42.6 \\
				\hline
				MRI & Head & 65.8\\
				\hline
				CT & Head & 135.7\\
				\hline
	\end{tabular}}}
\end{table}
\paragraph*{Quantitative analysis}
We evaluate the segmentation algorithm using the dataset by Bauer et al.~\citep{data/ICSFUS_2022, bauer2021generation}, which includes 52 subjects. For each subject, the dataset provides CT scans in both the inhale and exhale phases, as well as MR scans for the inhale and exhale phases. To assess the accuracy of the segmentation methods, we compared the segmentations produced for the same subject across the two imaging modalities, where the best result corresponds to two coincident surfaces. We used the Hausdorff distance and the mean of the distance distribution to quantify the similarity between surfaces. We perform this segmentation with both the graphic-based approach and the TotalSegmentator. In both cases, higher values indicate lower similarity and reduced segmentation accuracy. Fig.~\ref{fig:confrontoDistribuzioni} presents the results of the graphics-based segmentation and the TotalSegmentator. Owing to its general applicability across different imaging modalities, graphic-based segmentation achieves better results.
\begin{figure}[t]
	\centering
	\begin{tabular}{cccc}
		\includegraphics[width=0.7 \linewidth]{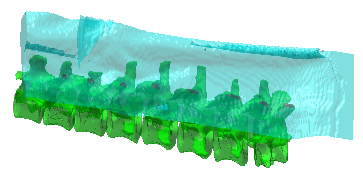}
	\end{tabular}
	\caption{\label{fig:landmarks}{Visualization support for surgical planning: the localisation of anatomical landmarks on the skin surface can be integrated into the trajectory planning method, to optimize surgical accuracy and guidance. The vertebral spine is shown in green, transverse processes in red, and the patient’s skin surface, extracted from CT data, in cyan. }}
\end{figure}
\section{Conclusions and future work\label{sec:CONCLUSION}}
We presented a comparative analysis of AI-based and algorithmic-based segmentation approaches, evaluating their strengths and limitations. Regarding the evaluation of skin segmentation methods for extracting a 3D surface representation of the body from anatomical images, the graphics-based approach proves to be versatile and adaptable, capable of segmenting various anatomical regions and working across different imaging modalities without requiring extensive training or large datasets. This aspect makes it particularly well-suited for applications involving multi-modal imaging. In contrast, TotalSegmentator effectively extracts the body surface while excluding the patient bed, but it is limited to CT images and relies on large, modality-specific datasets, which may present challenges in applications such as image fusion. Therefore, the choice of segmentation method should be guided by application-specific requirements and constraints.
\begin{figure}[t]
	\centering
	\begin{tabular}{cccc}
		(a)\includegraphics[height=0.2\linewidth]{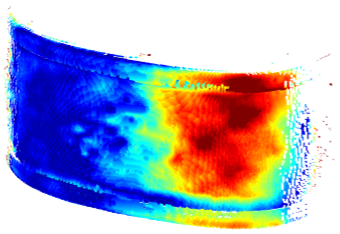}&
		(b)\includegraphics[height=0.2\linewidth]{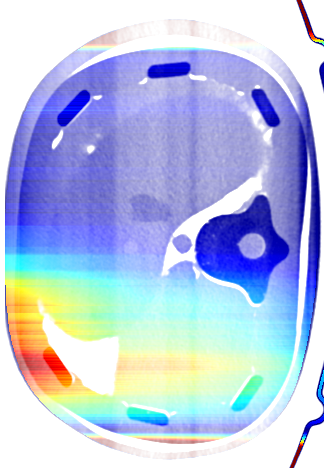}&
		(c)\includegraphics[height=0.15\linewidth]{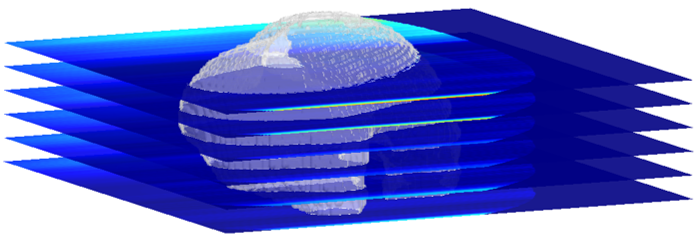}&
	\end{tabular}
	\caption{\label{fig:error_distribution}Coregistration error distance distribution on the surface (a). Corresponding error distribution inside the single slice (b). 3D Visualisation of the coregistration error distribution inside the volume, with a 3D model of the organ.}
\end{figure}
The 3D body model serves as both a clinical tool for intervention planning and an intuitive guide for users in understanding self-monitoring devices. Fig.~\ref{fig:landmarks} illustrates a 3D skin surface model that can be used to localise key spinal landmarks, such as spinous processes, providing valuable references for surgery, posture assessment, and monitoring. These landmarks can also be visually mapped onto the skin, allowing subject-specific identification from the outside and enhancing self-monitoring and user awareness. Additionally, visualising AI-generated results on 3D surfaces can improve the interpretability and transparency of black-box models, making them more acceptable in clinical settings. 

Future works will focus on applying 3D skin segmentation in surgical planning for uncertainty maps computation and visualisation, as they highlight areas of higher uncertainty in navigation and their proximity to critical organs (Fig.~\ref{fig:error_distribution}). Moreover, future work will focus on scoping the advancement of explainable and trustworthy AI methods for skin segmentation to address the challenges associated with the "black-box" nature of AI models. Additionally, Unsupervised Domain Adaptation techniques will be explored to enhance image acquisition flexibility in skin segmentation.
%\begin{acks}
%This work has been partially supported by the European Commission, NextGenerationEU, Missione 4 Componente 2, ``\emph{Dalla ricerca all’impresa}'', Innovation Ecosystem RAISE ``\emph{Robotics and AI for Socio-economic Empowerment}'', ECS00000035.
%\end{acks}
%
\subsection*{Acknowledgements}
M. Paccini and G. Patan\`e are part of the RAISE Innovation Ecosystem, funded by the European Union - NextGenerationEU and by the Ministry of University and Research (MUR), National Recovery and Resilience Plan (NRRP), Mission 4, Component 2, Investment 1.5, project ``RAISE - \emph{Robotics and AI for Socio-economic Empowerment}'' (ECS00000035). 

\bibliographystyle{elsarticle-harv}
{\small{\bibliography{Skin_segm_arxiv}}}

\begin{thebibliography}{40}
\expandafter\ifx\csname natexlab\endcsname\relax\def\natexlab#1{#1}\fi
\providecommand{\url}[1]{\texttt{#1}}
\providecommand{\href}[2]{#2}
\providecommand{\path}[1]{#1}
\providecommand{\DOIprefix}{doi:}
\providecommand{\ArXivprefix}{arXiv:}
\providecommand{\URLprefix}{URL: }
\providecommand{\Pubmedprefix}{pmid:}
\providecommand{\doi}[1]{\href{http://dx.doi.org/#1}{\path{#1}}}
\providecommand{\Pubmed}[1]{\href{pmid:#1}{\path{#1}}}
\providecommand{\bibinfo}[2]{#2}
\ifx\xfnm\relax \def\xfnm[#1]{\unskip,\space#1}\fi
%Type = Inproceedings
\bibitem[{Al-Faris et~al.(2014)Al-Faris, Ngah, Isa and Shuaib}]{al2014breast}
\bibinfo{author}{Al-Faris, A.Q.}, \bibinfo{author}{Ngah, U.K.},
  \bibinfo{author}{Isa, N.A.M.}, \bibinfo{author}{Shuaib, I.L.},
  \bibinfo{year}{2014}.
\newblock \bibinfo{title}{Breast {MRI} {Tumour} {Segmentation} {Using}
  {Modified} {Automatic} {Seeded} {Region} {Growing} {Based} on {Particle}
  {Swarm} {Optimization} {Image} {Clustering}}, in: \bibinfo{booktitle}{Soft
  Computing in Industrial Applications: Proceedings of the 17th Online World
  Conference on Soft Computing in Industrial Applications},
  \bibinfo{organization}{Springer}. pp. \bibinfo{pages}{49--60}.
%Type = Article
\bibitem[{Bauer et~al.(2021)Bauer, Russ, Waldkirch, T{\"o}nnes, Segars, Schad,
  Z{\"o}llner and Golla}]{bauer2021generation}
\bibinfo{author}{Bauer, D.F.}, \bibinfo{author}{Russ, T.},
  \bibinfo{author}{Waldkirch, B.I.}, \bibinfo{author}{T{\"o}nnes, C.},
  \bibinfo{author}{Segars, W.P.}, \bibinfo{author}{Schad, L.R.},
  \bibinfo{author}{Z{\"o}llner, F.G.}, \bibinfo{author}{Golla, A.K.},
  \bibinfo{year}{2021}.
\newblock \bibinfo{title}{Generation of annotated multimodal ground truth
  datasets for abdominal medical image registration}.
\newblock \bibinfo{journal}{International Journal of Computer Assisted
  Radiology and Surgery} \bibinfo{volume}{16}, \bibinfo{pages}{1277--1285}.
%Type = Article
\bibitem[{Baum et~al.(2012)Baum, Yap, Karampinos, Nardo, Kuo, Burghardt,
  Masharani, Schwartz, Li and Link}]{baum_does_2012}
\bibinfo{author}{Baum, T.}, \bibinfo{author}{Yap, S.P.},
  \bibinfo{author}{Karampinos, D.C.}, \bibinfo{author}{Nardo, L.},
  \bibinfo{author}{Kuo, D.}, \bibinfo{author}{Burghardt, A.J.},
  \bibinfo{author}{Masharani, U.B.}, \bibinfo{author}{Schwartz, A.V.},
  \bibinfo{author}{Li, X.}, \bibinfo{author}{Link, T.M.}, \bibinfo{year}{2012}.
\newblock \bibinfo{title}{Does vertebral bone marrow fat content correlate with
  abdominal adipose tissue, lumbar spine bone mineral density, and blood
  biomarkers in women with type 2 diabetes mellitus?}
\newblock \bibinfo{journal}{Journal of Magnetic Resonance Imaging}
  \bibinfo{volume}{35}, \bibinfo{pages}{117--124}.
%Type = Article
\bibitem[{Beare et~al.(2016)Beare, Yang, Maixner, Harvey, Kean, Anderson and
  Seal}]{beare_automated_2016}
\bibinfo{author}{Beare, R.}, \bibinfo{author}{Yang, J.Y.},
  \bibinfo{author}{Maixner, W.J.}, \bibinfo{author}{Harvey, A.S.},
  \bibinfo{author}{Kean, M.J.}, \bibinfo{author}{Anderson, V.A.},
  \bibinfo{author}{Seal, M.L.}, \bibinfo{year}{2016}.
\newblock \bibinfo{title}{Automated alignment of perioperative {{MRI}} scans:
  {A} technical note and application in pediatric epilepsy surgery}.
\newblock \bibinfo{journal}{Human Brain Mapping} \bibinfo{volume}{37},
  \bibinfo{pages}{3530--3543}.
%Type = Inproceedings
\bibitem[{Bi et~al.(2024)Bi, Pieper, Chlorogiannis, Golby and
  Frisken}]{bi2024open}
\bibinfo{author}{Bi, L.}, \bibinfo{author}{Pieper, S.},
  \bibinfo{author}{Chlorogiannis, D.D.}, \bibinfo{author}{Golby, A.J.},
  \bibinfo{author}{Frisken, S.}, \bibinfo{year}{2024}.
\newblock \bibinfo{title}{Open-source, deep-learning skin surface segmentation
  model for cost-effective neuronavigation accessible to low-resource
  settings}, in: \bibinfo{booktitle}{Medical Imaging 2024: Image Processing},
  \bibinfo{organization}{SPIE}. pp. \bibinfo{pages}{569--575}.
%Type = Inproceedings
\bibitem[{Bosnjak et~al.(2007)Bosnjak, Montilla, Villegas and
  Jara}]{bosnjak20073d}
\bibinfo{author}{Bosnjak, A.}, \bibinfo{author}{Montilla, G.},
  \bibinfo{author}{Villegas, R.}, \bibinfo{author}{Jara, I.},
  \bibinfo{year}{2007}.
\newblock \bibinfo{title}{{3D} {Segmentation} with an {Application} of {Level}
  {Set-method} using {MRI} {Volumes} for {Image} {Guided} {Surgery}}, in:
  \bibinfo{booktitle}{2007 29th Annual International Conference of the IEEE
  Engineering in Medicine and Biology Society}, \bibinfo{publisher}{IEEE}. pp.
  \bibinfo{pages}{5263--5266}.
%Type = Inproceedings
\bibitem[{Butz et~al.(2000)Butz, Warfield, Tuncali, Silverman, van Sonnenberg,
  Jolesz and Kikinis}]{butz2000pre}
\bibinfo{author}{Butz, T.}, \bibinfo{author}{Warfield, S.K.},
  \bibinfo{author}{Tuncali, K.}, \bibinfo{author}{Silverman, S.G.},
  \bibinfo{author}{van Sonnenberg, E.}, \bibinfo{author}{Jolesz, F.A.},
  \bibinfo{author}{Kikinis, R.}, \bibinfo{year}{2000}.
\newblock \bibinfo{title}{Pre-and {Intra-operative} {Planning} and {Simulation}
  of {Percutaneous} {Tumor} {Ablation}}, in: \bibinfo{booktitle}{Medical Image
  Computing and Computer-Assisted Intervention--MICCAI 2000: Third
  International Conference, Pittsburgh, PA, USA, October 11-14, 2000.
  Proceedings 3}, \bibinfo{organization}{Springer}. pp.
  \bibinfo{pages}{317--326}.
%Type = Article
\bibitem[{Caballo et~al.(2018)Caballo, Boone, Mann and
  Sechopoulos}]{caballo2018unsupervised}
\bibinfo{author}{Caballo, M.}, \bibinfo{author}{Boone, J.M.},
  \bibinfo{author}{Mann, R.}, \bibinfo{author}{Sechopoulos, I.},
  \bibinfo{year}{2018}.
\newblock \bibinfo{title}{An unsupervised automatic segmentation algorithm for
  breast tissue classification of dedicated breast computed tomography images}.
\newblock \bibinfo{journal}{Medical Physics} \bibinfo{volume}{45},
  \bibinfo{pages}{2542--2559}.
%Type = Article
\bibitem[{Chan and Vese(2001)}]{chan2001active}
\bibinfo{author}{Chan, T.F.}, \bibinfo{author}{Vese, L.A.},
  \bibinfo{year}{2001}.
\newblock \bibinfo{title}{Active contours without edges}.
\newblock \bibinfo{journal}{IEEE Transactions on Image Processing}
  \bibinfo{volume}{10}, \bibinfo{pages}{266--277}.
%Type = Article
\bibitem[{Diaz-Pinto et~al.(2024)Diaz-Pinto, Alle, Nath, Tang, Ihsani, Asad,
  Pérez-García, Mehta, Li, Flores, Roth, Vercauteren, Xu, Dogra, Ourselin,
  Feng and Cardoso}]{DIAZPINTO2024103207}
\bibinfo{author}{Diaz-Pinto, A.}, \bibinfo{author}{Alle, S.},
  \bibinfo{author}{Nath, V.}, \bibinfo{author}{Tang, Y.},
  \bibinfo{author}{Ihsani, A.}, \bibinfo{author}{Asad, M.},
  \bibinfo{author}{Pérez-García, F.}, \bibinfo{author}{Mehta, P.},
  \bibinfo{author}{Li, W.}, \bibinfo{author}{Flores, M.},
  \bibinfo{author}{Roth, H.R.}, \bibinfo{author}{Vercauteren, T.},
  \bibinfo{author}{Xu, D.}, \bibinfo{author}{Dogra, P.},
  \bibinfo{author}{Ourselin, S.}, \bibinfo{author}{Feng, A.},
  \bibinfo{author}{Cardoso, M.J.}, \bibinfo{year}{2024}.
\newblock \bibinfo{title}{Monai label: A framework for {AI}-assisted
  interactive labeling of {3D} medical images}.
\newblock \bibinfo{journal}{Medical Image Analysis} \bibinfo{volume}{95},
  \bibinfo{pages}{103207}.
%Type = Article
\bibitem[{Gillot et~al.(2022)Gillot, Baquero, Le, Deleat-Besson, Bianchi,
  Ruellas, Gurgel, Yatabe, Al~Turkestani, Najarian
  et~al.}]{gillot2022automatic}
\bibinfo{author}{Gillot, M.}, \bibinfo{author}{Baquero, B.},
  \bibinfo{author}{Le, C.}, \bibinfo{author}{Deleat-Besson, R.},
  \bibinfo{author}{Bianchi, J.}, \bibinfo{author}{Ruellas, A.},
  \bibinfo{author}{Gurgel, M.}, \bibinfo{author}{Yatabe, M.},
  \bibinfo{author}{Al~Turkestani, N.}, \bibinfo{author}{Najarian, K.}, et~al.,
  \bibinfo{year}{2022}.
\newblock \bibinfo{title}{Automatic multi-anatomical skull structure
  segmentation of cone-beam computed tomography scans using {3D} {UNETR}}.
\newblock \bibinfo{journal}{PLoS One} \bibinfo{volume}{17},
  \bibinfo{pages}{e0275033}.
%Type = Article
\bibitem[{Hirsch et~al.(2021)Hirsch, Huang and Parra}]{hirsch2021segmentation}
\bibinfo{author}{Hirsch, L.}, \bibinfo{author}{Huang, Y.},
  \bibinfo{author}{Parra, L.C.}, \bibinfo{year}{2021}.
\newblock \bibinfo{title}{Segmentation of {MRI} head anatomy using deep
  volumetric networks and multiple spatial priors}.
\newblock \bibinfo{journal}{Journal of Medical Imaging} \bibinfo{volume}{8},
  \bibinfo{pages}{034001--034001}.
%Type = Article
\bibitem[{Isensee et~al.(2021)Isensee, Jaeger, Kohl, Petersen and
  Maier-Hein}]{isensee2021nnu}
\bibinfo{author}{Isensee, F.}, \bibinfo{author}{Jaeger, P.F.},
  \bibinfo{author}{Kohl, S.A.}, \bibinfo{author}{Petersen, J.},
  \bibinfo{author}{Maier-Hein, K.H.}, \bibinfo{year}{2021}.
\newblock \bibinfo{title}{nn{U}-{Net}: a self-configuring method for deep
  learning-based biomedical image segmentation}.
\newblock \bibinfo{journal}{Nature Methods} \bibinfo{volume}{18},
  \bibinfo{pages}{203--211}.
%Type = Article
\bibitem[{Jermyn et~al.(2013)Jermyn, Ghadyani, Mastanduno, Turner, Davis,
  Dehghani and Pogue}]{jermyn_fast_2013}
\bibinfo{author}{Jermyn, M.}, \bibinfo{author}{Ghadyani, H.},
  \bibinfo{author}{Mastanduno, M.A.}, \bibinfo{author}{Turner, W.},
  \bibinfo{author}{Davis, S.C.}, \bibinfo{author}{Dehghani, H.},
  \bibinfo{author}{Pogue, B.W.}, \bibinfo{year}{2013}.
\newblock \bibinfo{title}{Fast segmentation and high-quality three-dimensional
  volume mesh creation from medical images for diffuse optical tomography}.
\newblock \bibinfo{journal}{Journal of Biomedical Optics} \bibinfo{volume}{18},
  \bibinfo{pages}{086007}.
%Type = Article
\bibitem[{Jing et~al.(2022)Jing, Liu, Wang, Zhang and Sun}]{jing2022recent}
\bibinfo{author}{Jing, J.}, \bibinfo{author}{Liu, S.}, \bibinfo{author}{Wang,
  G.}, \bibinfo{author}{Zhang, W.}, \bibinfo{author}{Sun, C.},
  \bibinfo{year}{2022}.
\newblock \bibinfo{title}{Recent advances on image edge detection: {A}
  comprehensive review}.
\newblock \bibinfo{journal}{Neurocomputing} \bibinfo{volume}{503},
  \bibinfo{pages}{259--271}.
%Type = Article
\bibitem[{Kamel~Boulos and Zhang(2021)}]{kamel2021digital}
\bibinfo{author}{Kamel~Boulos, M.N.}, \bibinfo{author}{Zhang, P.},
  \bibinfo{year}{2021}.
\newblock \bibinfo{title}{{Digital} {Twins}: {From} {Personalised} {Medicine}
  to {Precision} {Public} {Health}}.
\newblock \bibinfo{journal}{Journal of Personalized Medicine}
  \bibinfo{volume}{11}, \bibinfo{pages}{745}.
%Type = Article
\bibitem[{Khang et~al.(2022)Khang, Park, Lee, Kim, Song and
  Lee}]{khang2022computer}
\bibinfo{author}{Khang, S.}, \bibinfo{author}{Park, T.}, \bibinfo{author}{Lee,
  J.}, \bibinfo{author}{Kim, K.W.}, \bibinfo{author}{Song, H.},
  \bibinfo{author}{Lee, J.}, \bibinfo{year}{2022}.
\newblock \bibinfo{title}{Computer-{Aided} {Breast} {Surgery} {Framework}
  {Using} a {Markerless} {Augmented} {Reality} {Method}}.
\newblock \bibinfo{journal}{Diagnostics} \bibinfo{volume}{12},
  \bibinfo{pages}{3123}.
%Type = Article
\bibitem[{Koitka et~al.(2024a)Koitka, Baldini, Kroll, van Landeghem, Pollok,
  Haubold, Pelka, Kim, Kleesiek, Nensa et~al.}]{koitka2024saros}
\bibinfo{author}{Koitka, S.}, \bibinfo{author}{Baldini, G.},
  \bibinfo{author}{Kroll, L.}, \bibinfo{author}{van Landeghem, N.},
  \bibinfo{author}{Pollok, O.B.}, \bibinfo{author}{Haubold, J.},
  \bibinfo{author}{Pelka, O.}, \bibinfo{author}{Kim, M.},
  \bibinfo{author}{Kleesiek, J.}, \bibinfo{author}{Nensa, F.}, et~al.,
  \bibinfo{year}{2024}a.
\newblock \bibinfo{title}{{SAROS}: A dataset for whole-body region and organ
  segmentation in {CT} imaging}.
\newblock \bibinfo{journal}{Scientific Data} \bibinfo{volume}{11},
  \bibinfo{pages}{483}.
%Type = Article
\bibitem[{Koitka et~al.(2024b)Koitka, Baldini, Schmidt, Pollok, Pelka, Kohnke,
  Borys, Friedrich, Schaarschmidt, Forsting et~al.}]{koitka2024salt}
\bibinfo{author}{Koitka, S.}, \bibinfo{author}{Baldini, G.},
  \bibinfo{author}{Schmidt, C.S.}, \bibinfo{author}{Pollok, O.B.},
  \bibinfo{author}{Pelka, O.}, \bibinfo{author}{Kohnke, J.},
  \bibinfo{author}{Borys, K.}, \bibinfo{author}{Friedrich, C.M.},
  \bibinfo{author}{Schaarschmidt, B.M.}, \bibinfo{author}{Forsting, M.},
  et~al., \bibinfo{year}{2024}b.
\newblock \bibinfo{title}{{SALT}: {Introducing} a {Framework} for
  {Hierarchical} {Segmentations} in {Medical} {Imaging} using {Softmax} for
  {Arbitrary} {Label} {Trees}}.
\newblock \bibinfo{journal}{arXiv preprint arXiv:2407.08878} .
%Type = Article
\bibitem[{Le~Troter et~al.(2016)Le~Troter, Four{\'e}, Guye, Confort-Gouny,
  Mattei, Gondin, Salort-Campana and Bendahan}]{le2016volume}
\bibinfo{author}{Le~Troter, A.}, \bibinfo{author}{Four{\'e}, A.},
  \bibinfo{author}{Guye, M.}, \bibinfo{author}{Confort-Gouny, S.},
  \bibinfo{author}{Mattei, J.P.}, \bibinfo{author}{Gondin, J.},
  \bibinfo{author}{Salort-Campana, E.}, \bibinfo{author}{Bendahan, D.},
  \bibinfo{year}{2016}.
\newblock \bibinfo{title}{Volume measurements of individual muscles in human
  quadriceps femoris using atlas-based segmentation approaches}.
\newblock \bibinfo{journal}{Magnetic Resonance Materials in Physics, Biology
  and Medicine} \bibinfo{volume}{29}, \bibinfo{pages}{245--257}.
%Type = Article
\bibitem[{Lee et~al.(2018)Lee, Chang, Chang and Chang}]{lee_automated_2018}
\bibinfo{author}{Lee, C.Y.}, \bibinfo{author}{Chang, T.F.},
  \bibinfo{author}{Chang, N.Y.}, \bibinfo{author}{Chang, Y.C.},
  \bibinfo{year}{2018}.
\newblock \bibinfo{title}{An automated skin segmentation of {Breasts} in
  {Dynamic} {Contrast}-{Enhanced} {Magnetic} {Resonance} {Imaging}}.
\newblock \bibinfo{journal}{Scientific Reports} \bibinfo{volume}{8},
  \bibinfo{pages}{6159}.
%Type = Article
\bibitem[{Lenchik et~al.(2019)Lenchik, Heacock, Weaver, Boutin, Cook, Itri,
  Filippi, Gullapalli, Lee, Zagurovskaya, Retson, Godwin, Nicholson and
  Narayana}]{lenchik_automated_2019}
\bibinfo{author}{Lenchik, L.}, \bibinfo{author}{Heacock, L.},
  \bibinfo{author}{Weaver, A.A.}, \bibinfo{author}{Boutin, R.D.},
  \bibinfo{author}{Cook, T.S.}, \bibinfo{author}{Itri, J.},
  \bibinfo{author}{Filippi, C.G.}, \bibinfo{author}{Gullapalli, R.P.},
  \bibinfo{author}{Lee, J.}, \bibinfo{author}{Zagurovskaya, M.},
  \bibinfo{author}{Retson, T.}, \bibinfo{author}{Godwin, K.},
  \bibinfo{author}{Nicholson, J.}, \bibinfo{author}{Narayana, P.A.},
  \bibinfo{year}{2019}.
\newblock \bibinfo{title}{Automated {Segmentation} of {Tissues} {Using} {CT}
  and {{MRI}}: {A} {Systematic} {Review}}.
\newblock \bibinfo{journal}{Academic Radiology} \bibinfo{volume}{26},
  \bibinfo{pages}{1695--1706}.
%Type = Incollection
\bibitem[{Lorensen and Cline(1998)}]{lorensen1998marching}
\bibinfo{author}{Lorensen, W.E.}, \bibinfo{author}{Cline, H.E.},
  \bibinfo{year}{1998}.
\newblock \bibinfo{title}{Marching cubes: A high resolution {3D} surface
  construction algorithm}, in: \bibinfo{booktitle}{Seminal graphics: pioneering
  efforts that shaped the field}, pp. \bibinfo{pages}{347--353}.
%Type = Article
\bibitem[{Ma et~al.(2024)Ma, He, Li, Han, You and Wang}]{ma2024segment}
\bibinfo{author}{Ma, J.}, \bibinfo{author}{He, Y.}, \bibinfo{author}{Li, F.},
  \bibinfo{author}{Han, L.}, \bibinfo{author}{You, C.}, \bibinfo{author}{Wang,
  B.}, \bibinfo{year}{2024}.
\newblock \bibinfo{title}{Segment anything in medical images}.
\newblock \bibinfo{journal}{Nature Communications} \bibinfo{volume}{15},
  \bibinfo{pages}{654}.
%Type = Article
\bibitem[{Ognard et~al.(2019)Ognard, Mesrar, Benhoumich, Misery, Burdin and
  Ben~Salem}]{ognard_edge_2019}
\bibinfo{author}{Ognard, J.}, \bibinfo{author}{Mesrar, J.},
  \bibinfo{author}{Benhoumich, Y.}, \bibinfo{author}{Misery, L.},
  \bibinfo{author}{Burdin, V.}, \bibinfo{author}{Ben~Salem, D.},
  \bibinfo{year}{2019}.
\newblock \bibinfo{title}{Edge detector-based automatic segmentation of the
  skin layers and application to moisturization in high-resolution 3 {Tesla}
  magnetic resonance imaging}.
\newblock \bibinfo{journal}{Skin Research and Technology} \bibinfo{volume}{25},
  \bibinfo{pages}{339--346}.
%Type = Article
\bibitem[{Otsu(1979)}]{otsu1979threshold}
\bibinfo{author}{Otsu, N.}, \bibinfo{year}{1979}.
\newblock \bibinfo{title}{A {Threshold} {Selection} {Method} from
  {Gray}-{Level} {Histograms}}.
\newblock \bibinfo{journal}{IEEE Transactions on Systems, Man, and Cybernetics}
  \bibinfo{volume}{9}, \bibinfo{pages}{62--66}.
%Type = Article
\bibitem[{Paccini et~al.(2024)Paccini, Paschina, De~Beni, Stefanov, Kolev and
  Patan{\`e}}]{paccini2024us}
\bibinfo{author}{Paccini, M.}, \bibinfo{author}{Paschina, G.},
  \bibinfo{author}{De~Beni, S.}, \bibinfo{author}{Stefanov, A.},
  \bibinfo{author}{Kolev, V.}, \bibinfo{author}{Patan{\`e}, G.},
  \bibinfo{year}{2024}.
\newblock \bibinfo{title}{{US} \& {MR}/{CT} {Image} {Fusion} with {Markerless}
  {Skin} {Registration}: {A} {Proof} of {Concept}}.
\newblock \bibinfo{journal}{Journal of Imaging Informatics in Medicine}
  \bibinfo{volume}{38}, \bibinfo{pages}{1--14}.
%Type = Article
\bibitem[{Pan and Lu(2007)}]{pan2007bayes}
\bibinfo{author}{Pan, Z.}, \bibinfo{author}{Lu, J.}, \bibinfo{year}{2007}.
\newblock \bibinfo{title}{A {Bayes}-{Based} {Region}-{Growing} {Algorithm} for
  {Medical} {Image} {Segmentation}}.
\newblock \bibinfo{journal}{Computing in Science \& Engineering}
  \bibinfo{volume}{9}, \bibinfo{pages}{32--38}.
%Type = Article
\bibitem[{Rosset et~al.(2004)Rosset, Spadola and Ratib}]{rosset2004osirix}
\bibinfo{author}{Rosset, A.}, \bibinfo{author}{Spadola, L.},
  \bibinfo{author}{Ratib, O.}, \bibinfo{year}{2004}.
\newblock \bibinfo{title}{Osirix: an open-source software for navigating in
  multidimensional {DICOM} images}.
\newblock \bibinfo{journal}{Journal of Digital Imaging} \bibinfo{volume}{17},
  \bibinfo{pages}{205--216}.
%Type = Article
\bibitem[{Schindelin et~al.(2012)Schindelin, Arganda-Carreras, Frise, Kaynig,
  Longair, Pietzsch, Preibisch, Rueden, Saalfeld, Schmid
  et~al.}]{schindelin2012fiji}
\bibinfo{author}{Schindelin, J.}, \bibinfo{author}{Arganda-Carreras, I.},
  \bibinfo{author}{Frise, E.}, \bibinfo{author}{Kaynig, V.},
  \bibinfo{author}{Longair, M.}, \bibinfo{author}{Pietzsch, T.},
  \bibinfo{author}{Preibisch, S.}, \bibinfo{author}{Rueden, C.},
  \bibinfo{author}{Saalfeld, S.}, \bibinfo{author}{Schmid, B.}, et~al.,
  \bibinfo{year}{2012}.
\newblock \bibinfo{title}{Fiji: an open-source platform for biological-image
  analysis}.
\newblock \bibinfo{journal}{Nature Methods} \bibinfo{volume}{9},
  \bibinfo{pages}{676--682}.
%Type = Misc
\bibitem[{Scientific(2025)}]{amira}
\bibinfo{author}{Scientific, T.F.}, \bibinfo{year}{2025}.
\newblock \bibinfo{title}{Amira software for 3d visualization and analysis}.
\newblock \URLprefix \url{https://www.thermofisher.com/amira-avizo}.
  \bibinfo{note}{accessed: 2025-02-18}.
%Type = Article
\bibitem[{Scorza et~al.(2021)Scorza, El~Hadji, Cortes, Bertelsen, Cardinale,
  Baselli, Essert and De~Momi}]{scorza2021surgical}
\bibinfo{author}{Scorza, D.}, \bibinfo{author}{El~Hadji, S.},
  \bibinfo{author}{Cortes, C.}, \bibinfo{author}{Bertelsen, A.},
  \bibinfo{author}{Cardinale, F.}, \bibinfo{author}{Baselli, G.},
  \bibinfo{author}{Essert, C.}, \bibinfo{author}{De~Momi, E.},
  \bibinfo{year}{2021}.
\newblock \bibinfo{title}{Surgical planning assistance in keyhole and
  percutaneous surgery: {A} systematic review}.
\newblock \bibinfo{journal}{Medical Image Analysis} \bibinfo{volume}{67},
  \bibinfo{pages}{101820}.
%Type = Article
\bibitem[{Teng et~al.(2024)Teng, Li, Xin, Xiang, Huang, Zhou, Shi, Zhu, Cai,
  Peng et~al.}]{teng2024literature}
\bibinfo{author}{Teng, Z.}, \bibinfo{author}{Li, L.}, \bibinfo{author}{Xin,
  Z.}, \bibinfo{author}{Xiang, D.}, \bibinfo{author}{Huang, J.},
  \bibinfo{author}{Zhou, H.}, \bibinfo{author}{Shi, F.}, \bibinfo{author}{Zhu,
  W.}, \bibinfo{author}{Cai, J.}, \bibinfo{author}{Peng, T.}, et~al.,
  \bibinfo{year}{2024}.
\newblock \bibinfo{title}{A literature review of artificial intelligence ({AI})
  for medical image segmentation: from {AI} and explainable {AI} to trustworthy
  {AI}}.
\newblock \bibinfo{journal}{Quantitative Imaging in Medicine and Surgery}
  \bibinfo{volume}{14}, \bibinfo{pages}{9620}.
%Type = Article
\bibitem[{Teuwen et~al.(2021)Teuwen, Moriakov, Fedon, Caballo, Reiser, Bakic,
  Garc{\'\i}a, Diaz, Michielsen and Sechopoulos}]{teuwen2021deep}
\bibinfo{author}{Teuwen, J.}, \bibinfo{author}{Moriakov, N.},
  \bibinfo{author}{Fedon, C.}, \bibinfo{author}{Caballo, M.},
  \bibinfo{author}{Reiser, I.}, \bibinfo{author}{Bakic, P.},
  \bibinfo{author}{Garc{\'\i}a, E.}, \bibinfo{author}{Diaz, O.},
  \bibinfo{author}{Michielsen, K.}, \bibinfo{author}{Sechopoulos, I.},
  \bibinfo{year}{2021}.
\newblock \bibinfo{title}{Deep learning reconstruction of digital breast
  tomosynthesis images for accurate breast density and patient-specific
  radiation dose estimation}.
\newblock \bibinfo{journal}{Medical Image Analysis} \bibinfo{volume}{71},
  \bibinfo{pages}{102061}.
%Type = Incollection
\bibitem[{Tripathi and Rosak-Szyrocka(2025)}]{tripathi2025disruptive}
\bibinfo{author}{Tripathi, S.}, \bibinfo{author}{Rosak-Szyrocka, J.},
  \bibinfo{year}{2025}.
\newblock \bibinfo{title}{Disruptive {Innovation} in {Medical} {Image}
  {Segmentation}: {A} {Comparative} {Study} of {Traditional} and {AI}-{Based}
  {Approaches}}, in: \bibinfo{booktitle}{Impact of Artificial Intelligence on
  Society}. \bibinfo{publisher}{Chapman and Hall/CRC}, pp.
  \bibinfo{pages}{1--18}.
%Type = Inproceedings
\bibitem[{Wang et~al.(2012)Wang, Platel, Ivanovskaya, Harz and
  Hahn}]{wang_fully_2012}
\bibinfo{author}{Wang, L.}, \bibinfo{author}{Platel, B.},
  \bibinfo{author}{Ivanovskaya, T.}, \bibinfo{author}{Harz, M.},
  \bibinfo{author}{Hahn, H.K.}, \bibinfo{year}{2012}.
\newblock \bibinfo{title}{Fully automatic breast segmentation in {{3D}} breast
  {{MRI}}}, in: \bibinfo{booktitle}{2012 9th {IEEE} {International} {Symposium}
  on {Biomedical} {Imaging} ({ISBI})}, \bibinfo{publisher}{IEEE},
  \bibinfo{address}{Barcelona, Spain}. pp. \bibinfo{pages}{1024--1027}.
%Type = Article
\bibitem[{Wasserthal et~al.(2023)Wasserthal, Breit, Meyer, Pradella, Hinck,
  Sauter, Heye, Boll, Cyriac, Yang et~al.}]{wasserthal2023totalsegmentator}
\bibinfo{author}{Wasserthal, J.}, \bibinfo{author}{Breit, H.C.},
  \bibinfo{author}{Meyer, M.T.}, \bibinfo{author}{Pradella, M.},
  \bibinfo{author}{Hinck, D.}, \bibinfo{author}{Sauter, A.W.},
  \bibinfo{author}{Heye, T.}, \bibinfo{author}{Boll, D.T.},
  \bibinfo{author}{Cyriac, J.}, \bibinfo{author}{Yang, S.}, et~al.,
  \bibinfo{year}{2023}.
\newblock \bibinfo{title}{Totalsegmentator: {Robust} {Segmentation} of 104
  {Anatomic} {Structures} in {CT} {Images}}.
\newblock \bibinfo{journal}{Radiology: Artificial Intelligence}
  \bibinfo{volume}{5}.
%Type = Article
\bibitem[{Wolf et~al.(2005)Wolf, Vetter, Wegner, Nolden, Bottger, Hastenteufel,
  Kunert and Meinzer}]{wolf2005medical}
\bibinfo{author}{Wolf, I.}, \bibinfo{author}{Vetter, M.},
  \bibinfo{author}{Wegner, I.}, \bibinfo{author}{Nolden, M.},
  \bibinfo{author}{Bottger, T.}, \bibinfo{author}{Hastenteufel, M.},
  \bibinfo{author}{Kunert, T.}, \bibinfo{author}{Meinzer, H.P.},
  \bibinfo{year}{2005}.
\newblock \bibinfo{title}{The {Medical} {Imaging} {Interaction} {Toolkit}}.
\newblock \bibinfo{journal}{Medical Image Analysis} \bibinfo{volume}{9},
  \bibinfo{pages}{594--604}.
%Type = Article
\bibitem[{Yushkevich et~al.(2006)Yushkevich, Piven, Hazlett, Smith, Ho, Gee and
  Gerig}]{yushkevich2006user}
\bibinfo{author}{Yushkevich, P.A.}, \bibinfo{author}{Piven, J.},
  \bibinfo{author}{Hazlett, H.C.}, \bibinfo{author}{Smith, R.G.},
  \bibinfo{author}{Ho, S.}, \bibinfo{author}{Gee, J.C.},
  \bibinfo{author}{Gerig, G.}, \bibinfo{year}{2006}.
\newblock \bibinfo{title}{User-guided {3D} active contour segmentation of
  anatomical structures: significantly improved efficiency and reliability}.
\newblock \bibinfo{journal}{Neuroimage} \bibinfo{volume}{31},
  \bibinfo{pages}{1116--1128}.
%Type = Misc
\bibitem[{Z{\"o}llner(2022)}]{data/ICSFUS_2022}
\bibinfo{author}{Z{\"o}llner, F.}, \bibinfo{year}{2022}.
\newblock \bibinfo{title}{{Multimodal ground truth datasets for abdominal
  medical image registration [data]}}.
\newblock \DOIprefix\doi{10.11588/data/ICSFUS}.

\end{thebibliography}
\end{document}